\documentclass[12pt]{article}
 \usepackage[dvips]{graphicx}
 \usepackage[dvips]{graphics}
 \usepackage{amsmath}
 \usepackage{enumerate}
 \usepackage{psfrag}
 \newcommand{\newc}{\newcommand}
 \newc{\ra}{\rightarrow}
 \newc{\lra}{\leftrightarrow}
 \newc{\beq}{\begin{equation}}
 \newc{\eeq}{\end{equation}}

\begin{document}

 \begin{center}
 {\LARGE \bf Axisymmetric equilibria  with
 anisotropic resistivity and toroidal flow }
 \vspace{2mm} \\
 {\large G. Poulipoulis$^\dag$\footnote{me00584@cc.uoi.gr}, G. N. Throumoulopoulos$^\dag$\footnote{gthroum@cc.uoi.gr},
 H. Tasso$^\star$\footnote{het@ipp.mpg.de}}
 \vspace*{1mm} \\
 $^\dag${\it University of Ioannina, Association Euratom - Hellenic
 Republic,
 \vspace{-1mm} \\
 Section of Theoretical Physics, GR 451 10 Ioannina, Greece}
 \vspace{1mm} \\

 \noindent $^\star${\it  Max-Planck-Institut f\"{u}r
 Plasmaphysik, Euratom Association,
 \vspace{-1mm} \\
 D-85748 Garching, Germany }
 \end{center}

 \begin{center}
 {\bf Abstract}
 \end{center}

 The  equilibrium   of an axisymmetric
 magnetically confined plasma with anisotropic resistivity and
 toroidal flow is investigated in the framework of magnetohydrodynamics (MHD).
 The stationary states
 are determined by an    elliptic differential
 equation for the poloidal magnetic flux function $\psi$,
 a Bernoulli equation for the pressure and two  relations
 for the resistivity components $\eta_\parallel$ and $\eta_\perp$ parallel and perpendicular
 to the magnetic field. The flow can affect the equilibrium
 properties solely  in the presence of  toroidicity because in the limit
 of infinite aspect ratio the axial velocity does not appear in the equilibrium equations.
 The equilibrium characteristics of  a tokamak
 with  rectangular cross-section are studied by means of
 eigenfunctions in connection with exact
 solutions  for the cases of
 ``compressible" flows with constant temperature, $T(\psi)$, but
 varying density on magnetic surfaces
 and incompressible ones with constant density, $\varrho(\psi)$, but
 varying temperature thereon.
 Those eigenfunctions can describe either single or multiple toroidal
 configurations. In the former case the equilibrium has the following characteristics:
 (i) the $\eta_\parallel$- and $\eta_\perp$-profiles on the poloidal cross-section
 having a minimum close to the magnetic axis, taking large
 values on the boundary and satisfying the relation $\eta_\perp > \eta_\parallel$
 are roughly collisional (ii) the electric field
 perpendicular to the magnetic surfaces  possesses two local extrema within the plasma and
 vanishes on the boundary and (iii) the toroidal current density  is  peaked close to the magnetic axis
 and vanishes on the boundary.
 The impact of the flow and the aspect ratio on the aforementioned
 quantities is evaluated for both ``compressible" and incompressible flows.
 \vspace{0.4cm} \\

 \newpage
 \noindent
 {\large \bf 1. Introduction}\\ 

 Understanding the equilibrium properties of a magnetically confined
  plasma is one of the key issues in the
 fusion research. The majority of  equilibrium studies up to date concern plasmas of zero electrical
 resistivity and have been based
 on the Grad-Schl\"uter-Shafranov equation which describes the magnetohydrodynamic (MHD) equilibria of
 axisymmetric plasmas. Additional  physical input in terms of  resistivity and plasma
 flow, however, are very important. In addition to its apparent role for
 Ohmic  heating,  the importance of resistivity  is connected to the operation of a
 steady-state fusion reactor which will involve  time scales much longer than
 the resistive-MHD one. Also an attractive feature of such a reactor would be  operation under minimum
 sources of mass, momentum and energy. Sheared flow  plays a key role in the formation of
 both edge transport barriers (L-H transition) \cite{Ter,deG}  and internal transport barriers
 \cite{Wan}-\cite{Co} in tokamaks. This flow is associated  with sheared
 electric field profiles \cite{KHB}-\cite{Mei} with significant
 amplitudes in the barrier region. Another pertinent
 quantity is the safety factor  \cite{Jo}-\cite{Qui} closely related to the toroidal current density.

 Theoretically it was proved long time ago  \cite{Ta} that axisymmetric
 resistive MHD equilibria with scalar resistivity uniform on magnetic surfaces  are not compatible
 with the Grad-Schl\"uter-Shafranov equation and  the poloidal current density should vanish.
 The nonexistence of axisymmetric equilibria with constant resistivity
 was also suggested in Refs. \cite{Mo1,Mo2}.
 To examine whether these undesirable properties can be removed by including, in addition to resistivity
 in Ohm's law, flow
 and viscosity terms in the momentum equation is a formidable task,
 which should be preceded by a step by step thorough investigation
 of particular cases  possibly introducing additional physical input
 each time. In this respect two of the authors studied axisymmetric
 equilibria with scalar resistivity and flow purely toroidal
 \cite{Th1} and  parallel to the magnetic field {\bf B} \cite{Th2}
 and found that the flows considered can not remove the
 aforementioned inconsistencies.  Non-vanishing poloidal currents are
 possible in steady states with parallel flows  in the presence of
 anisotropy, i.e. for
 different resistivity components $\eta_\parallel$ and $\eta_\perp$
 parallel and perpendicular to $\bf B$ \cite{Th3};
 however, in this case  neither $\eta_\parallel$ nor $\eta_\perp$ can
 be uniform on magnetic surfaces. The sole external source
 in Refs. \cite{Th1}-\cite{Th3}
 is  the toroidal current loop voltage. Resistive equilibria
 in a similar spirit of minimum external sources were investigated in Refs. \cite{Mo1,Mo2} and
 \cite{BaLe}-\cite{KaMo2}.
 The particular flow directions considered  in Refs. \cite{Th1}-\cite{Th3}
 are not inconsistent with Pfisrch-Schl\"uter diffusion in the sense
 that the equilibrium solutions constructed therein
 neither exclude nor can be included
 in possible Pfirsch-Schl\"uter-diffusion solutions (having velocity
 components perpendicular to the magnetic surfaces). Also, it is
 reminded that the pertinent  pioneering study \cite{PfSc}
 does not include external sources of current and neglects the  flow term in the momentum equation.

 In this report we extend the studies \cite{Th1}-\cite{Th3} to equilibria
 with toroidal flows and anisotropic resistivity. There are two
 advantageous features of toroidal flows compared
 with  parallel ones: (i) they are associated with  electric fields ${\bf E}_p$ perpendicular
 to the magnetic surfaces and (ii) as we will show exact solutions with magnetic surfaces of uniform temperature,
 $T=T(\psi)$,  are
 possible. Equilibria with incompressible flows having uniform density but varying temperature
 on magnetic surfaces will also be examined.
 In this respect it is noted that although purely toroidal axisymmetric
 flows are inherently incompressible
 because of symmetry, $T(\psi)$-equilibria can be regarded as ``compressible" in the sense
 that the density varies on magnetic surfaces.
 In both cases the  study can be carried out analytically up to the construction of exact solutions.
 In particular   equilibrium eigenstates  of a tokamak with  rectangular cross-section will be derived
 in connection with exact solutions describing either single toroidal or
 multitoroidal configurations. Furthermore for single toroidal eigenstates we will study the characteristics
 of the conductivity components
 $\sigma_\parallel$ and $\sigma_\perp$, the electric field ${\bf E}_p$, and the toroidal current density $J_\phi$
 along with the impact of the flow on these quantities. This impact will be examined by varying
 a sound-speed Mach number $M_0$ for ``compressible" flows and a parameter $A$ relating to the $\varrho$-
 and ${\bf E}_p$- profiles and their variation perpendicular to the magnetic surfaces for incompressible ones.
 In particular the uniformity of
 $\sigma_\parallel$ and $\sigma_\perp$ on magnetic surfaces will be examined
 independently of solutions. Also it will be
 shown that the impact of the flow on the equilibrium is crucially related to the toroidicity.

 An outline of the report is as follows. Reduced equilibrium equations involving $\psi$, the pressure
 and the resistivity components are derived in Sec. 2. Tokamak eigenstates
 are constructed in Sec. 3 for both ``compressible" and incompressible flows. The equilibrium characteristics of
 $\sigma_\perp$, $\sigma_\parallel$, ${\bf E}_p$, and $J_\phi$ along
 with the impact of the flow and the aspect ratio on them is the subject of Sec. 4. The possible role of the
 flow shear is also briefly discussed therein. The conclusions are summarized in Sec. 5. \vspace{0.4cm} \\

 \noindent
 {\large \bf 2. Equilibrium equations}\\

 In this section we shall derive reduced equilibrium equations for
 an axisymmetric magnetically confined plasma with anisotropic resistivity and toroidal flow.
 The procedure is unified in the sense that
 relevant energy equations or equations of state are not adopted from the beginning;
 they will specified when necessary later.

 The starting equations in standard notation and convenient units are the following:
 \begin{eqnarray}
 \mbox{\boldmath $\nabla$}\cdot (\varrho {\bf v})=0, \label{1} \\
 \varrho ({\bf v}\cdot \mbox{\boldmath
 $\nabla$}){\bf v}={\bf J}\times{\bf B}-\mbox{\boldmath $\nabla$}P,
 \label{2} \\
 \mbox{\boldmath $\nabla$}\times{{\bf E}}=0, \label{3} \\
 \mbox{\boldmath $\nabla$}\times{\bf B}={\bf J}, \label{4} \\
 \mbox{\boldmath $\nabla$}\cdot{\bf B}=0, \label{5} \\
 {{\bf E}}+{\bf v}\times{\bf B}=\mbox{\boldmath $\eta$}\cdot{\bf J}=
 \eta_\parallel\cdot{\bf J}_\parallel+\eta_\perp{\bf J}_\perp, \label{6} \\
 \mbox{ An energy equation or equation of state }, \label{7}
 \end{eqnarray}
 where
 \beq
 \mbox{\boldmath $\eta$}=\left(
 \begin{array}{cc}
 \eta_\parallel & 0 \\
 0 & \eta_\perp
 \end{array} \right)
 \label{8}
 \eeq
 is the  resistivity tensor;  the indices $\parallel$ and $\perp$
 indicate directions parallel and perpendicular to $\bf B$;
 accordingly ${\bf J_\parallel}=({\bf J\cdot b}){\bf b}$, ${\bf
 J_\perp}={\bf b}\times ({\bf J}\times{\bf b})={\bf J}-{\bf J}_\parallel$
 with ${\bf b}={\bf B}/B$.
 The procedure to follow is based on identifying some integrals as flux functions,
 i.e. functions constant on magnetic surfaces, and  reducing the set of Eqs. (\ref{1}-\ref{6})
 by projecting the momentum equation (\ref{2}) and Ohm's law (\ref{6})
 along the toroidal direction, the poloidal one (or parallel to $\bf B$ when convenient),
 and perpendicular to the magnetic surfaces.
 Important information is also drawn from an integral
 form of (\ref{6}) [Eq. (\ref{13}) below].

 In cylindrical coordinates $(R,\phi ,z)$ with $z$
 corresponding to the axis of symmetry  the equilibrium quantities  for the case under consideration
 do not depend on the toroidal angle $\phi$; the
 toroidal velocity and the divergence-free magnetic field and current
 density can be expressed,  with the aid of Amp$\acute{e}$re's law, in terms of
 the functions $K(R,z)$, $\psi(R,z)$ and $I(R,z)$ as:
 \begin{eqnarray}
 \varrho{\bf v}=K\mbox{\boldmath $\nabla$}\phi , \label{9} \\
 {\bf B}=I\mbox{\boldmath $\nabla$}\phi +\mbox{\boldmath $\nabla$}\phi\times \mbox{\boldmath $\nabla$}\psi
 , \label{10} \\
 {\bf J}=\Delta^*\psi \mbox{\boldmath $\nabla$}\phi -\mbox{\boldmath $\nabla$}\phi\times
 \mbox{\boldmath $\nabla$}I, \label{11}
 \end{eqnarray}
 where $\psi$ labels the magnetic surfaces and $\Delta^*\equiv R^2$ {\boldmath $\nabla$}
 $\cdot$({\boldmath $\nabla$}$/R^2)$.

 By projecting the momentum equation along the toroidal direction
 one obtains
 \beq
 \mbox{\boldmath $\nabla$}\phi\cdot (\mbox{\boldmath $\nabla$}\psi\times \mbox{\boldmath $\nabla$}I)=0,
 \label{12}
 \eeq
 which implies  that $I=I(\psi )$. Therefor, unlike the case of parallel flow \cite{Th3},  the current
 surfaces coincide with the magnetic ones irrespective of
 equation of state.
 Integration of  (\ref{6}) along a contour $c$
 defined by the cut of an arbitrary current surface with the
 poloidal plane yields the equation:
 \beq
 \int_c{{{\bf E}}\cdot d{\bf l}}+\int_c{({\bf v}\times{\bf B})\cdot
 d{\bf l}}=\int_c({\mbox{\boldmath $\eta$}\cdot{\bf J})\cdot d{\bf l}},
 \label{13}
 \eeq
 where $d{\bf l}=\mbox{\boldmath $\nabla$} \phi\times \mbox{\boldmath $\nabla$}
 \psi/|\mbox{\boldmath $\nabla$}\phi\times \mbox{\boldmath $\nabla$} \psi|$
 is the unit vector along the poloidal direction.
 Since in equilibrium it holds that $\partial{\bf B}/\partial t=0$,
 the first integral on the left-hand side of (\ref{13}) vanishes by
 Stoke's theorem. Also the second integral vanishes due to the
 toroidal direction of the flow. For the integral on the right-hand side
 to vanish the integrand must necessarily do so because the ${\bf J}_{pol}$-lines are
 closed, nested and $\mbox{\boldmath $\nabla$} \cdot {\bf J}=0$;
 therefor it should hold locally
 \beq
 (\mbox{\boldmath $\eta$}\cdot{\bf J})\cdot d{\bf l}=(\mbox{\boldmath $\eta$}\cdot{\bf J})_{pol}=0.
 \label{13a}
 \eeq
 For isotropic resistivity, $\eta_\perp=\eta_\parallel$, (\ref{13a}) implies that the
 poloidal current density must vanish. In the presence of  anisotropy, however,
 non-vanishing poloidal current densities are possible  as expected because the toroidal
 electric field can drive a current in the poloidal direction. The rest of the report concerns
 equilibria with non-vanishing poloidal current densities.

 Expressing the electric field on the poloidal cross-section in terms of the electrostatic
 potential, ${\bf E}_p=-\mbox{\boldmath $\nabla$} \Phi$, the component of
 local Ohm's law (\ref{6}) in the poloidal direction on account of (\ref{13a}) yields
 \beq
 \mbox{\boldmath $\nabla$}\phi\cdot(\mbox{\boldmath $\nabla$} \Phi\times\mbox{\boldmath $\nabla$}\psi)=0.
 \label{14}
 \eeq
 which implies that $\Phi=\Phi(\psi)$; therefor  ${\bf E}_p$ is perpendicular
 to the magnetic surfaces. The total electric field is given by
 $$
 {\bf E}= V_c\mbox{\boldmath $\nabla$} \phi + {\bf E}_p = V_c\mbox{\boldmath $\nabla$}
 \phi - \Phi^\prime\mbox{\boldmath $\nabla$} \psi,
 $$
 where $2\pi V_c$ is the constant toroidal loop voltage and the prime denotes derivative
 with respect to $\psi$.
 Subsequently, the component of (\ref{6}) along
 $\mbox{\boldmath $\nabla$}\psi$ yields
 \beq
 (\Phi^\prime-\frac{K}{\varrho R^2} )\cdot\left|\mbox{\boldmath
 $\nabla$}\psi\right|^2 =0,
 \label{16}
 \eeq
 and therefor the quantity
 \beq
 \frac{K}{\varrho R^2}\equiv\omega= \Phi^\prime,
 \label{17}
 \eeq
 identified as the rotation frequency, is a flux function
 $\omega=\omega (\psi)$.
 Eq. (\ref{13a}) and the component of (\ref{6}) in the  toroidal direction respectively  yield
 the following equations:
 \begin{eqnarray}
 -\frac{\Delta\eta}{(BR)^2}(I
 \Delta^*\psi-I'|\mbox{\boldmath $\nabla$}\psi|^2)-
 \eta_\perp I'=0, \label{20} \\
 V_c=\Delta\eta\frac{I}{(BR)^2}[I'|\mbox{\boldmath
 $\nabla$}\psi|^2-I\Delta^*\psi]+\eta_\perp\Delta^*\psi, \label{21}
 \end{eqnarray}
 where $\Delta\eta=\eta_\perp-\eta_\parallel$.  Any equilibrium solution should be compatible
 with (\ref{20}) and (\ref{21}) which, accordingly, can be solved
 for $\eta_\perp$ and $\eta_\parallel$ to yield
 \begin{eqnarray}
 \eta_\perp=\frac{V_c}{\Delta^*\psi+II'}, \label{22} \\
 \eta_\parallel=\eta_\perp\Big(1+\frac{I'(BR)^2}{I\Delta^*\psi
 -I'|\mbox{\boldmath $\nabla$}\psi|^2}\Big). \label{23}
 \end{eqnarray}

 With the aid of the integrals $I=I(\psi)$, $\Phi=\Phi(\psi)$ and $\omega=\omega (\psi)$ the components of
 (\ref{2}) along ${\bf B}$ and {\boldmath $\nabla$}$\psi$ respectively yield
 \begin{eqnarray}
 \bigg[\frac{\mbox{\boldmath $\nabla$}P}{\varrho}-\mbox{\boldmath
 $\nabla$}\Big(\frac{\omega^2R^2}{2}\Big)\bigg]\cdot{\bf B}=0,
 \label{18} \\
 \big[\Delta^*\psi+II'\big]| \mbox{\boldmath
 $\nabla$}\psi|^2+R^2\bigg[ \mbox{\boldmath
 $\nabla$}P-\varrho\omega^2 \mbox{\boldmath
 $\nabla$}\Big(\frac{R^2}{2}\Big)\bigg]\cdot\mbox{\boldmath $\nabla$}\psi=0.
 \label{19}
 \end{eqnarray}
 Owing to the axisymmetry and the toroidal direction of the flow these equations
 do not contain the resistivity and are identical in form with the respective ideal-MHD
 equations.

 In order to reduce Eqs. (\ref{18}) and (\ref{19}) further an energy
 equation or equation of state is necessary. Owing to the large heat
 conduction along $\bf B$,  isothermal magnetic surfaces, $T=T(\psi)$, is an
 appropriate equation of state for fusion plasmas. In this case
 employing the ideal gas law, $P=\lambda\varrho T$, integration
 of  (\ref{18}) yields
 \beq
 P=P_s(\psi)\exp{\Big(\frac{\omega^2R^2}{2\lambda T}\Big)}
 \label{24}
 \eeq
 where $P_s(\psi)$ is the pressure in the absence of flow.
 With the aid of (\ref{24}), Eq. (\ref{19}) leads to the final ``compressible" equation
 \beq
 \Delta^*\psi+II'+R^2\bigg[P_s'+P_s\frac{R^2}{2}\Big(\frac{\omega^2}{\lambda
 T}\Big)'\bigg]\exp{\Big(\frac{\omega^2R^2}{2\lambda T}\Big)}=0.
 \label{25}
 \eeq
 For ideal plasmas (\mbox{\boldmath $\eta$}={\bf 0}) Eq. (\ref{25})
 was originally obtained in Ref. \cite{MaPe}.

 An alternative equation of state is incompressibility:
 \beq
 \mbox{\boldmath $\nabla$}\cdot{\bf v}=0.
 \label{26}
 \eeq
 Consequently, (\ref{1}) implies that the density is a flux function,
 $\varrho=\varrho(\psi)$, and therefor one can find along the same lines the following incompressible equations
 for $P$ and $\psi$:
 \beq
 P=P_s(\psi)+\frac{R^2\varrho\omega^2}{2}.
 \label{27}
 \eeq
 \beq
 \Delta^*\psi+II'+R^2P_s'+\frac{R^4}{2}(\varrho\omega^2)'=0.
 \label{28}
 \eeq
 Eq. (\ref{28}) is  identical with  a particular  form of the
 axisymmetric equilibrium equation for incompressible flow of
 arbitrary direction obtained in Ref. \cite{TaTh98} for ideal
 plasmas.

 Once Eqs. (\ref{25}) and (\ref{28}) are solved for $\psi$ the resistivity components can be
 determined by (\ref{22}) and (\ref{23}). In general inspection of Eqs. (\ref{22}) and (\ref{23})
 implies, like the case of parallel flows \cite{Th3}, that neither $\eta_\perp$ nor $\eta_\parallel$ can be
 uniform on magnetic surfaces; indeed  solving (\ref{25}) and
 (\ref{28}) for $\Delta^\star\psi$ and substituting into (\ref{22})
 and (\ref{23}), one can see that $\eta_\perp$ and $\eta_\parallel$
 depend, in addition to $\psi$, explicitly on $R$ (and on $|\nabla
 \psi|^2$ as concerns $\eta_\parallel$).
 However as we will see in Sec. 4, $\eta_\perp$ and $\eta_\parallel$
 can be collisional-like, viz. they can have a minimum close to the magnetic axis,
 take very large values on the boundary and it holds  that $\eta_\perp>\eta_\parallel$.
 Reasons for  temperature deviations  on magnetic surfaces, which can result to
 non uniformity of $\eta_\parallel$ and $\eta_\perp$ thereon, are discussed
 in Sec. III of Ref. \cite{Th3}.

 Summarizing this section, the MHD equilibrium states of an
 axisymmetric plasma with anisotropic resistivity and toroidal flow
 is governed by an elliptic differential equation for the poloidal
 magnetic flux function [Eq. (\ref{25}) for
 ``compressible" flow and (\ref{28})  for incompressible one],
 a Bernoulli relation for the pressure  and self-consistent expressions
 for the resistivities $\eta_\parallel$ and $\eta_\perp$.
 Both Eqs. (\ref{25}) and (\ref{28}) contain four flux-functions,
 three out of which, i.e. $P_s$, $I$ and $\omega$, are common. The fourth function is
 $T$ for the ``compressible equation'' and $\varrho$ for the incompressible
 one. For vanishing flow  (\ref{25}) and (\ref{28})  reduce to the Grad-Schl\"uter-Shafranov
 equation.\vspace{0.4cm} \\

 \noindent
 {\large \bf 3. Exact solutions}\\

 Linearized forms of Eqs. (\ref{25}) and (\ref{28}) in connection with  appropriate
 assignments of the free flux functions they contain can be solved analytically.
 In the present study we will employ exact solutions  as follows.\\

 \noindent
 {\em ``Compressible'' flow}\\
 The ansatz used to linearize Eq. (\ref{25}) is \cite{ClFa} \cite{ThPa}
 \begin{eqnarray}
 I^2=I_0^2+I_1^2\psi^2 \nonumber \\
 P_s=2P_0\psi^2 \label{ans:1} \\
 \frac{\omega^2}{\lambda T}=\frac{\gamma M_0^2}{R_0^2}=\mbox{constant} \nonumber
 \end{eqnarray}
 Here, $I_0/R$ is the toroidal vacuum field, the parameter $I_1$
 describes the magnetic properties of the plasma;  $P_0$, $\gamma$,
 and $M_0$ are a pressure parameter, the ratio of  specific heats,
 and the Mach number with respect to the  sound-speed at a reference
 point ($z=0, R=R_0$) with $R_0$ to be specified later. Note that the
 toroidal current density profile  can vanish on the plasma  boundary
 via  (\ref{10}).

 Eq. (\ref{25}) then
 has a separable solution, ${\cal R}(R){\cal Z}(z)$, when the
 constant of separation is equal to $R_0I_1$. For configurations
 symmetric with respect to mid-plane $z=0$ this solution is
 written in the form
 \beq
 \psi (x,y)=C_1\bigg[J_0\Big(\frac{2\, \tau{\sqrt{e^
 {\frac{\gamma\,{M_0}^2\,x^2}{2}}}}}{\gamma\, {M_0}^2}\Big)+C_2Y_0\Big(\frac{2\, \tau{\sqrt{e^
 {\frac{\gamma\,{M_0}^2\,x^2}{2}}}}}{\gamma\,
 {M_0}^2}\Big)\bigg]\cos{(R_0I_1 y)},
 \label{29}
 \eeq
 where
 $x=R/R_0$ and $y=z/R_0$; $J_0$ and $Y_0$ are
 zeroth-order Bessel functions of first- and second-kind
 respectively; and
 $\tau^2 \equiv 4P_0R_0^4$.\\

 \noindent
 {\em Incompressible flow}\\
 In this case the ansatz employed to linearize (\ref{28}) is \cite{ThPo}
 \begin{eqnarray}
 I^2=I_0^2+I_1^2\psi^2 \nonumber \\
 P_s=2P_0\psi^2 \label{ans:2} \\
 \left(\varrho \omega^2\right)^\prime =\Big[\frac{K^2}{\varrho R^4}\Big]'=2A\psi \nonumber
 \end{eqnarray}
 The third of  equations (\ref{ans:2}) on account of (\ref{17}) indicates that $A$ is associated
 with the density and electric field profiles and their variation (shear) perpendicular to the
 magnetic surfaces [$(\varrho\omega^2)^\prime \neq 0$]. The polarity of ${\bf E}_p$ and the aforementioned
 shear permits $A$ to take either positive or negative values. This
 is a remarkable difference as compared with  the choice (\ref{ans:1}) which is shearless
 [$(\omega^2/\lambda T)^\prime = 0$].
 Also, note that, unlike $M_0$ in (\ref{ans:1}), $A$ is dimensional.

 A separable solution is now expressed in terms of the first- and
 second-kind Airy functions, $A_i$ and $B_i$, as \cite{ThPo}
 \begin{eqnarray}
 \psi (x,y)=C_1\Bigg[
 Ai\bigg(\Big(\frac{AR_0}{4}\Big)^{-2/3}\Big(\frac{AR_0^6}{4}x^2-P_1R_0^4\Big)\bigg) \nonumber \\
 +C_2 Bi\bigg(\Big(\frac{AR_0}{4}\Big)^{-2/3}\Big(\frac{AR_0^6}{4}x^2-P_1R_0^4\Big)\bigg)\Bigg]
 \cos{(R_0I_1 y)}.
 \label{30}
 \end{eqnarray}

\begin{figure}[!htb]
 \begin{center}
 \includegraphics[scale=0.8]{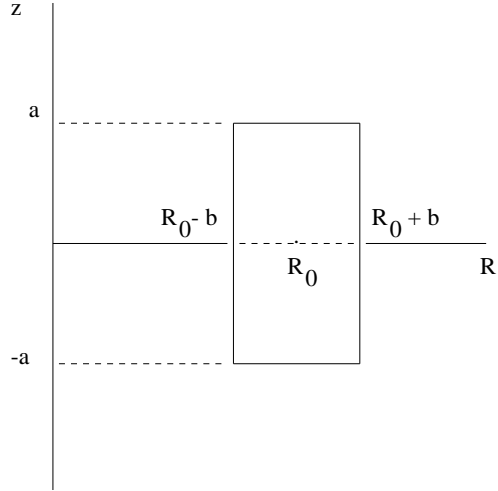}
 \caption{The  cross-section of the plasma boundary.
 The aspect ratio $\alpha$ is defined as $R_0/b$, where $R_0$ is the
 geometric center of the configuration.}
 \label{fig:1}
 \end{center}
 \end{figure}

 In connection with solutions (\ref{29}) and (\ref{30}) we are interested in the steady
 states of a tokamak the plasma of which is bounded by a
 conducting wall of rectangular cross-section, as shown in Fig.
 \ref{fig:1}.
 In addition, we assume that the plasma boundary coincides with
 the outermost magnetic surface. Thus, the magnetic field is tangential to
 and the pressure $P$ must vanish  on the boundary \cite{com2}; accordingly, the function $\psi$
 should satisfy the following boundary conditions
 \beq
 \psi(y_\pm) = 0
 \label{25a}
 \eeq
 and
 \beq
 \psi(x_\pm) = 0,
 \label{25b}
 \eeq
 where $y_\pm=\pm a/R_0$ and $x_\pm=1\pm b/R_0$. The equilibrium becomes
 then a boundary-value problem. Eigenstates can be determined by
 imposing conditions (\ref{25a}) and (\ref{25b}) directly to solutions
 (\ref{29}) and (\ref{30}).
 Specifically, (\ref{25a}) applied to  the $z$-dependent part of the solutions
 yields the eigenvalues
 \beq
 I_1^\ell = \frac{1}{\mbox{a}}\left(\ell \pi -\frac{\pi}{2}\right),\ \
 \ell=1,2,\ldots
 \label{25c}
 \eeq
 for the quantity $I_1$ which is related to the poloidal current function
 $I(\psi)$.
 The respective eigenfunctions are associated with
 configurations possessing $\ell$ magnetic axes parallel to the axis of symmetry.
 Condition (\ref{25b}) is pertinent to the $R$-dependent part of
 the solution. Owing to the flow  this part contains the parameter
 $M_0$ in the ``compressible'' case and $A$ in the incompressible
 one in addition to the  pressure parameter $P_0$. Thus,  condition
 (\ref{25b}) can determine flow eigenvalues   depending on the parameter
 $P_0$ which remains free, $F^n(P_0)$ ($n=1,2,3,\ldots$) with F standing for either  $M_0$ or $A$, or vice
 versa, pressure eigenvalues $P_0^n(F)$ with $F$ being free.
 The other parameters $C_1$ and $C_2$ contained in (\ref{29})
 and (\ref{30}) are adapted to normalize $\psi$ with respect
 to the magnetic axis and to satisfy the boundary condition
 (\ref{25b}) respectively. The eigenfunctions in association with $F^n(P_0)$ [or $P_0^n(F)$] are connected to
 configurations having $n$ magnetic axes perpendicular to the axis of symmetry.
 Therefor the total equilibrium eigenfunctions $\psi_{\ell n}={\cal Z}_\ell(z){\cal R}_n(R)$
 describe multitoroidal configurations  having $\ell\times n$
 magnetic axes.

 On the basis of the above solutions one can
 evaluate the impact of the flow  on the
 resistivity components $\eta_\perp$ [Eq. (\ref{22})] and $\eta_\parallel$
 [Eq. (\ref{23})], the  electric field perpendicular to the magnetic surfaces [${\bf E}_p=-\mbox{
 \boldmath $\nabla$}\Phi=-\Phi'\mbox{\boldmath $\nabla$}\psi$] and  the toroidal current density $J_\phi$ [Eq.
 (\ref{11})] for both ``compressible" and incompressible flows.
 It is emphasized that this impact
 is crucially related to the toroidicity because in the limit of
 infinite aspect ratio the equilibrium equations do not contain the
 axial velocity regardless of``compressibility". Indeed,
 for a cylindrical plasma of arbitrary cross-section the equations
 respective to (\ref{18}) and (\ref{19}) read
 \begin{eqnarray}
 {\bf B}\cdot\mbox{\boldmath $\nabla$}P=0
 \label{18:b} \\
 \nabla^2\psi+\Big(P+\frac{B_z^2}{2}\Big)'=0.
 \label{19:b}
 \end{eqnarray}
 For ideal plasmas Eqs. (\ref{18:b}) and (\ref{19:b}) follow respectively from (16) and (17)
 of Ref \cite{ThTa97} for vanishing poloidal velocity
 ($F^\prime=0$ therein).
 Therefor the flow may have an impact on equilibrium only in the
 presence of toroidicity.
 Also, note that the pressure becomes a flux-function. Because of the importance of toroidicity
 the impact of the aspect ratio on equilibrium in addition to that of the flow  will be
 evaluated in the next section. \vspace{0.4cm} \\

 \noindent
 {\large \bf 4. Impact of the flow and aspect ratio on equilibrium}\\

\begin{figure}[!h]
 \begin{center}
 \psfrag{z}{$z$}
 \psfrag{R}{$R$}
 \includegraphics[scale=0.6, trim=40 46 41 31 clip]{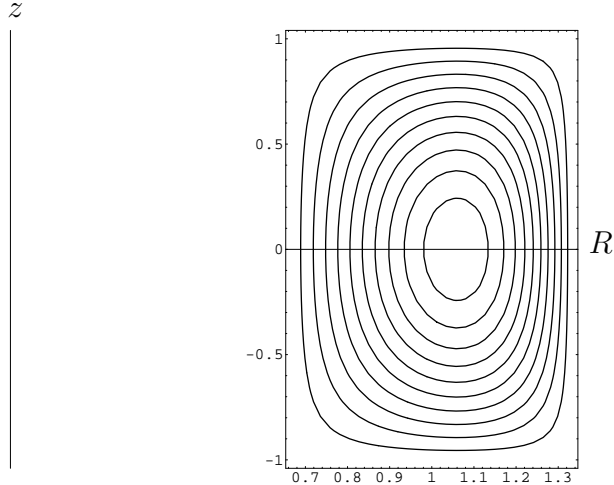}
 \caption{Magnetic surface cross-sections for the single toroidal configuration
 in connection with the eigenfunction $\psi_{11}$ for $M_0=0.5$ and $\alpha=3$.}
 \label{fig:2}
 \end{center}
 \end{figure}

\begin{figure}[!htb]
 \begin{center}
 \psfrag{sv}{$\frac{\sigma_\perp}{\sigma_c}$}
 \psfrag{x}{$x$}
 \psfrag{a}{a)}
 \psfrag{b}{b)}
 \psfrag{c}{c)}
 \psfrag{d}{d)}
 \includegraphics[scale=0.8]{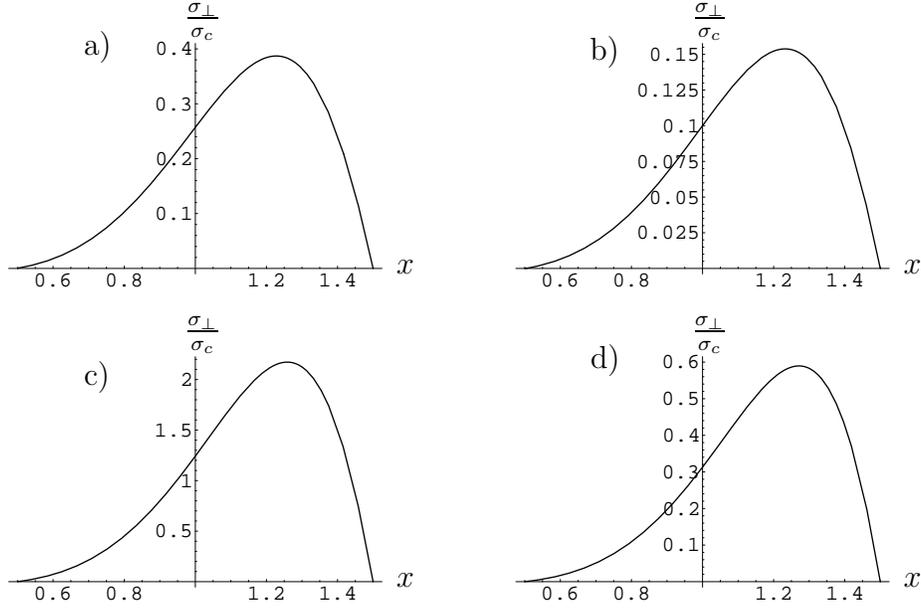}
 \caption{ A set of curves  demonstrating the oscillation of the
 profile of   $\sigma_\perp$ on the mid-plane $z=0$, normalized with respect to a constant
 value $\sigma_c$,  for ``compressible" flow when the value of the function $\psi$
 on the magnetic axis  is flow dependent  and  the Mach number $M_0$
 increases:   a) $M_0=0.1$,
 b) $M_0=0.2$, c) $M_0=0.5$ and d) $M_0=0.6$. The aspect ratio is $\alpha=2$;
  $x\equiv R/R_0$ with the   vertical axis being placed
  at the position of the geometric center ($x=1$).}
 \label{fig:3}
 \end{center}
 \end{figure}

 The quantities to be examined are  the  conductivity components  $\sigma_\perp=1/\eta_\perp$
 and $\sigma_\parallel=1/\eta_\parallel$,
 the electric field ${\bf E}_p$  and  the toroidal current density $J_\phi$.
 Both ``compressible"  and incompressible flows will be studied for single
 toroidal configurations. The study is based on
 the eigenfunction $\psi_{11}$ which for ``compressible" flow is shown
 in  Fig. \ref{fig:2} and will be made by varying the flow
 parameters $M_0$ and $A$ for ``compressible" and incompressible flows
 respectively.
 For any value of $M_0$ or $A$ the respective lowest
 eigenvalue of the pressure parameter $P_0$ will be calculated numerically. The variation
 of $M_0$ and $A$  will correspond to the same range of eigenvalues of  $P_0$. Specifically,
 for aspect ratio $\alpha=3$ the flow
 parameters $M_0$ and $A$  will  be ranged in the intervals [0.1, 0.7]
 and [-0.001, -0.01] respectively. For $\alpha=2$ the respective intervals will be
 [0.1, 1] and [-0.001, -0.08] unless stated otherwise.
 Variation of the flow parameters in connection with  the results to be presented will
 refer to these intervals. The tokamak scaling $B_p \approx 0.1 B_\phi$
 will be used for the calculations. Also, we should make the following
 clarification: it occurs that  solutions (\ref{29}) and (\ref{30}) oscillate as the flow
 parameters are varied, viz. for any given point $(R,z)$ these
 solutions considered as functions $\psi(M_0)$and $\psi(A)$
 take successively larger and lower values as $M_0$ and $A$ are varied
 monotonically. This, would give rise
 to an oscillatory behavior to all physical quantities which
 is physically unjustifiable. For the conductivity components
 this  can be seen in Fig. \ref{fig:3}.
 To avoid this difficulty the solutions
 will be normalized in such a way that the poloidal magnetic flux on
 the magnetic axis is unity irrespective of flow. This
 is accomplished by choosing appropriately the parameter $C_1$ for each
 value of $M_0$ or $A$ [$C_1(M_0)$ for ``compressible" flow and $C_1(A)$ for
 ``incompressible" one]. Also, solutions (\ref{29}) and (\ref{30}) have
 a strong parametric dependence on the flow parameters [note the exponential dependence
 of (\ref{29}) on $M_0^2$]. Consequently, this dependence results in large quantitative changes
 in the physical quantities for large $M_0$ ($M_0 \approx 1$) or
 small $A$ ($A\approx -0.01$) which most probably overestimate the impact of
 flow. In addition, it is noted that except for the conductivity
 components an increase of $M_0$ has qualitatively the same impact on the
 physical quantities to be examined with that caused by a decrease of $A$.

 The results concerning the  characteristics of the quantities  $\sigma_\perp$, $\sigma_\parallel$,
 ${\bf E}_p$, and
 $J_\phi$ and the impact of the flow and  $\alpha$ on them  are as follows.\vspace{0.1cm} \\

 \noindent
 {\em Conductivity components}

 \begin{enumerate}
 \item For both ``compressible" and incompressible flows the profiles of $\sigma_\perp$ and
  $\sigma_\parallel$ on the poloidal cross-section
 are  collisional-like, i.e. they have a maximum
 close to the magnetic axis vanish on the boundary and stands in most
 of the cases that $\sigma_\parallel
 >\sigma_\perp$\cite{com3}.
 Profiles of the resistivity components on the mid-plane $z=0$  are shown in Fig. \ref{fig:4}
 for ``compressible" flows and in Fig.  \ref{fig:4a} for incompressible ones.
 \begin{figure}[!htb]
 \begin{center}
 \psfrag{sv}{$\frac{\sigma_\perp}{\sigma_c}$}
 \psfrag{sp}{$\frac{\sigma_\parallel}{\sigma_c}$}
 \psfrag{x}{$x$}
 \psfrag{zo}{$M_0=0.1$}
 \psfrag{zs}{$M_0=0.7$}
 \includegraphics[scale=1]{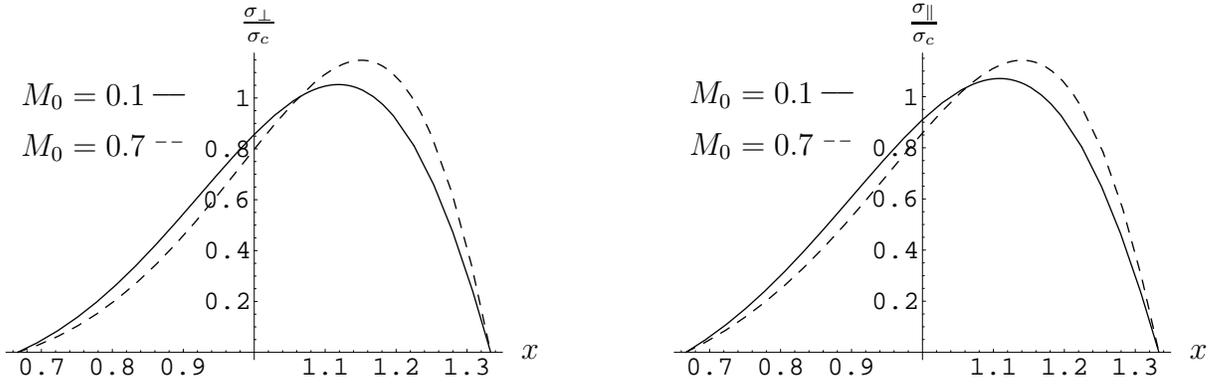}
 \caption{The figure shows  the increase of the maximum
 of $\sigma_\perp$ and $\sigma_\parallel$ and the displacement of its
 position outwards for ``compressible"  flow due to the increase of the  Mach-number $M_0$.}
 \label{fig:4}
 \end{center}
 \end{figure}
 \begin{figure}[!htb]
 \begin{center}
 \psfrag{x}{$x$}
 \psfrag{sv}{$\frac{\sigma_\perp}{\sigma_c}$}
 \psfrag{sp}{$\frac{\sigma_\parallel}{\sigma_c}$}
 \psfrag{mzzo}{$A=-0.001$}
 \psfrag{mzo}{$A=-0.01$}
 \includegraphics[scale=1]{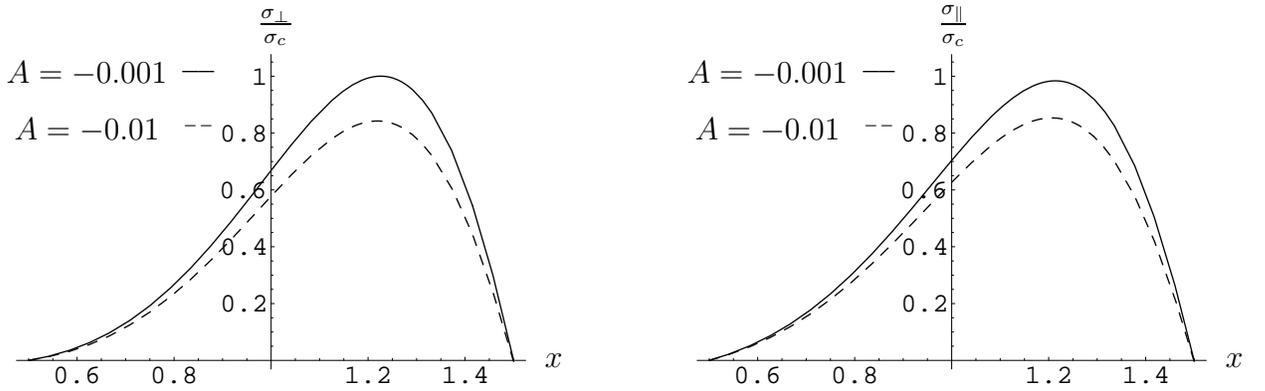}
 \caption{The figure shows  the decrease of the maximum
 of $\sigma_\perp$ and $\sigma_\parallel$ but the insensitivity of its
 position for  incompressible flow as the  parameter $A$ decreases. }
 \label{fig:4a}
 \end{center}
 \end{figure}
 For vanishing flow and $\alpha=3$
 one finds $\Delta \sigma /\sigma_\parallel \equiv(\sigma_\parallel-\sigma_\perp)/\sigma_\parallel=0.2122$;
 for the reversed-field-pinch scaling, $B_\phi\approx B_p$, the value of
 $(\sigma_\parallel-\sigma_\perp)/\sigma_\parallel$ becomes double, i.e. 0.4244.
  Also, for $\alpha=3$ increase of
 $M_0$ from  0.1 to 0.7 results in a percentage decrease of $\Delta \sigma$ by 4\% while decrease
 of $A$ from -0.001 to -0.01 leads to an increase of $\Delta \sigma$ by 3.4\%.
 \item The maximum of the profiles takes larger values as $M_0$ increases and lower values as $A$ decreases.
 For $\alpha=3$ increase of  $M_0$ (from 0.1 to 0.7) leads to percentage increases of
 $\sigma_\perp$ and $\sigma_\parallel$ by 9\% and 7\% respectively. For incompressible flows the respective
 decreases are 73\% and 67\% (for $\alpha=3$ and decrease of $A$
 from -0.001 to -0.01).
 \item The position of the maximum is shifted outwards from the axis of symmetry as $M_0$
 increases and is nearly insensitive to the variation of $A$ \cite{com1}
 (see Figs. \ref{fig:4} and \ref{fig:4a}). For example
 for $\alpha=3$ the  position of the maximum of $\sigma_\perp$ changes from 1.119 to 1.151 (when $M_0$ varies
 from 0.1 to 0.7).
 \item The lower $\alpha$
 \begin{itemize}
                     \item the larger the decrease of $\Delta\sigma=\sigma_\parallel-\sigma_\perp$  as $M_0$
                     increases;
                     \item the smaller the increase of $\Delta \sigma$ as $A$ decreases;
                     \item the larger the increase of the $\sigma_\perp$- and $\sigma_\parallel$-maximum
                     as $M_0$ increases;
                     \item the smaller the decrease of the $\sigma_\perp$- and $\sigma_\parallel$-maximum
                     as $A$ decreases;
                     \item the larger the shift of the position of $\sigma_\perp$- and $\sigma_\parallel$-maximum
                      as $M_0$ increases.
 \end{itemize}
 Those   conclusions  become evident if
 the following  results for $\alpha=2$ are compared with those
 presented above for $\alpha=3$:

 \begin{itemize}
      \item  the decrease of $\Delta \sigma$ for compressible flow is 13\%;
       \item the increase of $\Delta \sigma$ for incompressible flow is
       1\%. The respective variation of $A$ in this case is
       [-0.001,-0.01] (see \cite{com3});
      \item the increase of the conductivity maximum values for ``compressible" flows are  44\% ($\sigma_\perp$)
      and 27\% ($\sigma_\parallel$);
      \item the decrease of the conductivity maximum values for incompressible flows are  43.7\% ($\sigma_\perp$)
      and 63\% ($\sigma_\parallel$);
      \item the position of the maximum of $\sigma_\perp$ for ``compressible" flow is displaced
      from 1.228 to 1.33.
 \end{itemize}
 \end{enumerate}
 The ``compressibility" dependent impact of the flow on $\sigma_\perp$, $\sigma_\parallel$
 and $\Delta\sigma$ may relate to the
 flow shear of the incompressible solution (\ref{30}) while the ``compressible" solution (\ref{29})
 is shearless. \vspace{0.2cm} \\

 \noindent
 {\em Electric field}

\begin{figure}[!htb]
 \begin{center}
 \psfrag{x}{$x$}
 \psfrag{Er}{$\frac{E_r}{E_c}$}
 \psfrag{zo}{$M_0=0.1$}
 \psfrag{ze}{$M_0=0.8$}
 \includegraphics[scale=0.8]{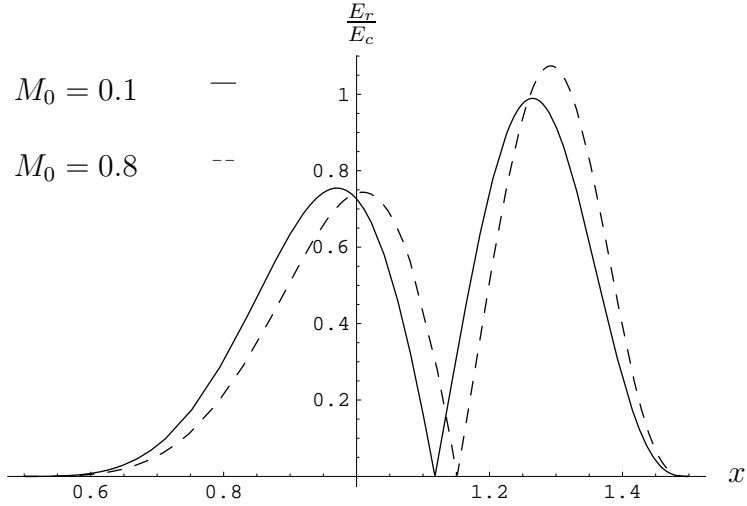}
 \caption{Electric field profiles on the mid-plane $z=0$ for ``compressible" flow
 Comparison of the two curves indicates the increase of the local maxima of ${\bf E}_p$
 and the outward shift of the their positions  as the
 Mach number $M_0$ increases from 0.8 to 1. The point in between the two maxima at
 which ${\bf E}_p=0$ corresponds to the magnetic axis. The aspect ratio is $\alpha=2$
 and the value of the shaping parameter $n$ [Eq. (\ref{omeg})] is 3.}
 \label{fig:5}
 \end{center}
 \end{figure}

\begin{figure}[!h]
 \begin{center}
 \psfrag{x}{$x$}
 \psfrag{Er}{$\frac{E_r}{E_c}$}
 \psfrag{no}{$n=1$}
 \psfrag{nt}{$n=3$}
 \includegraphics[scale=0.8]{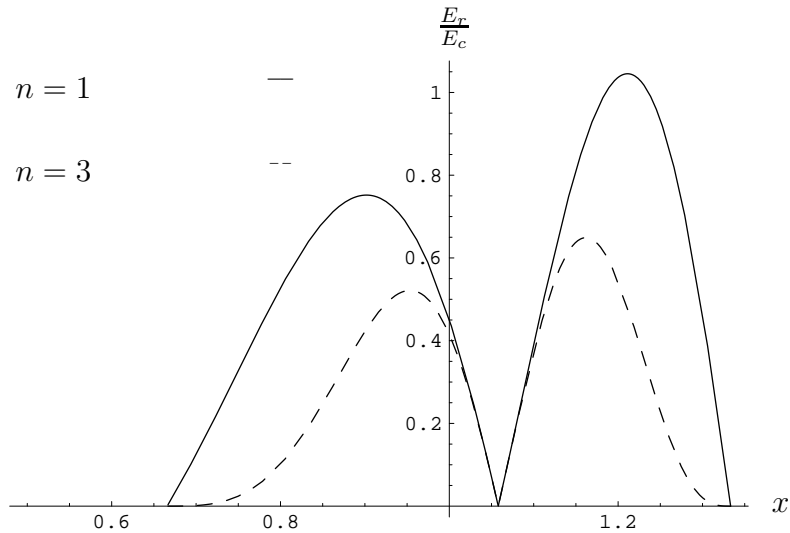}
 \caption{Electric field profiles on the mid-plane $z=0$ for $\alpha=3$, $M_0=0.4$
 and two different values of the shaping parameter $n$: $n=1$ and $n=3$.}
 \label{fig:6}
 \end{center}
 \end{figure}

 Prescribing the rotation frequency $\omega(\psi)$ by
 \beq
 \omega=\omega_0\psi^n,
 \label{omeg}
 \eeq
 where $n$ is a shaping parameter, the  electric field, ${\bf
 E}_p=-\Phi^\prime\mbox{\boldmath $\nabla$} \psi$,  with the help of
 (\ref{17}) becomes
 \beq
 {\bf E}_p=-\omega_0\psi^n\mbox{\boldmath $\nabla$}\psi.
 \label{33}
 \eeq
 The profile of ${\bf E}_p$ on the poloidal cross-section possesses two
 local extrema within the plasma volume  one on the left-hand-side and
 the other on the right-hand-side of the magnetic axis with opposite sign,
 which means that the electric filed has different polarity [as can also be deduced
 from Eq. (\ref{33})] and is similar to ones
 observed in equilibria with internal transport barriers.
 The profile $|E_r|$, on the  mid-plane $y=0$ is shown in Fig. \ref{fig:5}.
 Increase of  $M_0$ or decrease of $A$ leads to an increase of both
 local maximum values of $|E_r|$ with the one outwards more than the one
 inwards the magnetic axis and to a displacement of their positions
 outwards. For ``compressible" flows this is shown in (Fig. \ref{fig:5}).
 For $\alpha=3$ the maximum values of $|E_r|$ increase by 4\% for
 ``compressible" flow (increase of $M_0$ from 0.1 to 0.7) and 5.6\% for
 incompressible one (decrease of $A$ from -0.001 to -0.01). The lower $\alpha$
 the larger the increase of the maxima and the displacement of their positions
 outwards. As an example, for $\alpha=2$ the increase of the maximum $|E_r|$
 becomes as large as 15\% for ``compressible" flow with strong increase occurring
 above $M_0=0.8$ and 9\% for incompressible flow.
 Also, when the parameter $n$ takes larger values the maxima of $|E_r|$ take lower values
 but the profile of $E_r$ becomes steeper and more localized (Fig. \ref{fig:6});
 therefor the shear $ S_{E_r}=\partial E_r/\partial x $ is also increased.

 \noindent
 {\em The toroidal current density}

\begin{figure}[!h]
 \begin{center}
 \psfrag{x}{$x$}
 \psfrag{jz}{$\frac{J_\phi}{(J_\phi)_c}$}
 \psfrag{zo}{$M_0=0.1$}
 \psfrag{ze}{$M_0=0.8$}
 \includegraphics[scale=0.8]{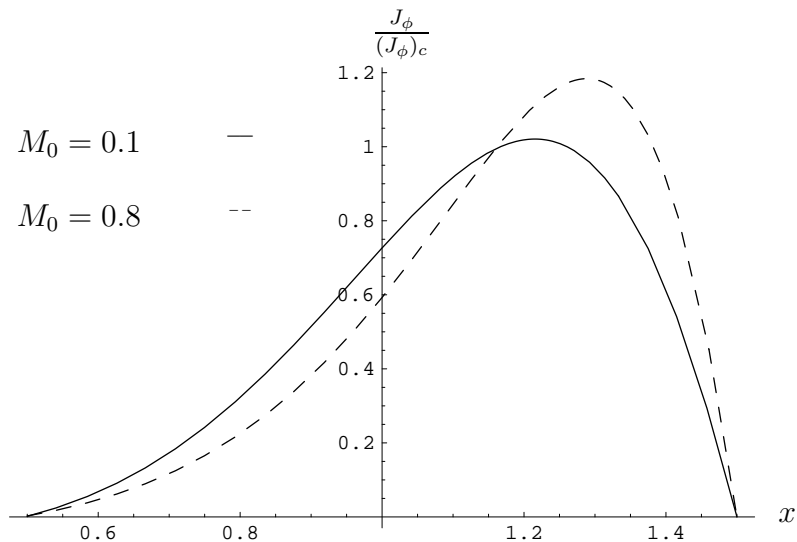}
 \caption{Toroidal current density profiles on the mid-plane $z=0$ for ``compressible" flows
 Comparison of the two curves indicates the increase of the maximum and the outward displacement of its
 position  as the Mach number  $M_0$ increases from 0.1 to 0.8. The aspect
 ratio is $\alpha =2$.}
 \label{fig:7}
 \end{center}
 \end{figure}

 The profile of the toroidal current density,
 \beq
 {J_\phi}=\frac{1}{R}{ \Delta}^*\psi,
 \label{34}
 \eeq
 is peaked with its maximum in the vicinity of the magnetic axis  and vanishes on the plasma boundary
 (Fig. \ref{fig:7}).
 Like ${\bf E}_p$, as $M_0$ increases or $A$ decreases the maximum of $J_\phi$
 takes larger values and its position is displaced outwards. Also, the impact of
 the variation of the flow parameters is stronger as the aspect ratio becomes smaller.
 In particular for $\alpha=3$ and $\alpha=2$ the maximum value of the current density  increases
 by 8\% and 35.5\% respectively  for compressible flows (increase of $M_0$). The respective
 increases for incompressible flows (decrease of $A$) are 5.6\% and 18\%. \vspace{0.4cm} \\

 \noindent
 {\large \bf 5. Conclusions}\\

 In this report we have investigated the MHD equilibrium of an axisymmetric magnetically
 confined plasma with anisotropic resistivity and toroidal flow subjected to the  single
 external source of toroidal current loop voltage by including
 the flow term in the momentum equation. Equilibria
 of this kind are inherently free of Pfirsch-Schl\"uter diffusion. Also, unlike the case
 of parallel flows,  there is an  electric field ${\bf E}_p$ perpendicular to the magnetic surfaces.
 The study includes ``compressible" flows with varying density
 but uniform temperature on magnetic surfaces and incompressible ones with uniform density but  varying temperature
 thereon. It turns out that the equilibrium states are  determined  by an elliptic
 differential equation for the poloidal magnetic flux function
 $\psi$ [Eq. (\ref{25}) for compressible flows  and (\ref{28}) for incompressible one],
 a Bernoulli equation for the pressure
 [respective Eqs. (\ref{24}) and (\ref{27})] and two relations for the
 resistivities $\eta_\perp$ [Eq. (\ref{22})] and  $\eta_\parallel$ [Eq. (\ref{23})] parallel and perpendicular
 to the magnetic field. Owing to axisymmetry and the toroidal direction of the flow,  the equilibrium equations
 and pressure relations are identical in form with the respective ideal-MHD ones. The impact of the flow on
 equilibrium can be ``activated" solely in the presence of toroidicity because the cylindrical equilibrium equations
 do not contain the axial velocity.

 The  equilibrium of  a tokamak plasma  bounded by a rectangular cross-section
 has been studied by means of equilibrium eigenfunctions in connection with  exact solutions
 for ``compressible"   and incompressible flows. These eigenfunctions can describe either single toroidal
 or multiple toroidal configurations. In the former case we have studied  the  characteristics of
 the conductivities $\sigma_\perp$ and  $\sigma_\parallel$, the electric  field ${\bf E}_p$,
 and the toroidal current density $J_\phi$ as well as  how they are affected by the flow. The impact of the flow
 has been examined by varying pertinent flow parameters, i.e. a sound-speed Mach number $M_0$
 for the ``compressible" flow and a parameter $A$ relating  to
 the density and electric field and their variations perpendicular to the magnetic surfaces for the
 incompressible one. For (i) single toroidal configurations (ii) eigenfunctions normalized so that  $\psi$ is
 unity on the magnetic axis regardless of flow  (a normalization made to avoid physically
 unjustifiable oscillation of the solutions on the
 flow parameters) and (iii) variation of the flow parameters corresponding to the same variation of
 the lowest eigenvalue for the pressure parameter  $P_0$  we came to the following conclusions:
 \begin{enumerate}
  \item   For a  toroidal frequency  $\omega(\psi)$ peaked on the
         magnetic axis and vanishing on the boundary [Eq. (\ref{omeg})] the
         profile of $|{\bf E}_p|$ on the poloidal cross-section
         has two maxima located the one on the left-hand-side
         and the other on the right-hand-side of the magnetic axis
         and it vanishes on the  boundary. When the maximum of  $\omega$ becomes larger and its
         profile more localized,
         the profile of $|{\bf E}_p|$ becomes as well more localized  though its maxima become smaller.

   \item The profile of $J_\phi $ is peaked with its maximum close to the magnetic
         axis and vanishes on the boundary.

   \item Although the conductivity components can not be uniform on magnetic surfaces
         (this follows in general  by inspection of the equilibrium equations)
         their profiles are roughly collisional, viz. they have a maximum in the
         vicinity of the magnetic axis, vanish on the boundary and it holds, in
         most of the cases that $\sigma_\parallel>\sigma_\perp$.

 \item As $M_0$ increases or $A$ decreases the local maxima of ${\bf E}_p$ and  $J_\phi$
        take larger values and their positions are shifted outwards from the axis of symmetry.

 \item The impact of the variation of the flow parameters on $\sigma_\perp$ and $\sigma_\parallel$ rely on
        ``compressibility": like ${\bf E}_p$ and $J_\phi$, the maxima of $\sigma_\perp$ and $\sigma_\parallel$
        become larger as $M_0$ increases and their positions are displaced outwards but these maxima
        become smaller and their positions are nearly not affected as $A$ decreases. Also, the larger
        the $M_0$ the smaller $\sigma_\parallel-\sigma_\perp$ but the smaller $A$ the larger
        $\sigma_\parallel-\sigma_\perp$.

 \item For  a given value of $M_0$, the lower the aspect ratio $\alpha$ the
       smaller the maxima of $\sigma_\parallel$, $\sigma_\perp$ and $J_\phi$
       but the larger the maximum of ${\bf E}_p$.

 \item For  a given value of $A$, the lower the aspect ratio $\alpha$ the
       larger the maxima of $\sigma_\parallel$, $\sigma_\perp$ and ${\bf E}_p$
       but the smaller the maximum of $J_\phi$.

 \item For increase of $M_0$ or decrease of $A$  (corresponding
       to the same variation of the lowest eigenvalue of $P_0$), the lower $\alpha$
       the higher the variation of the maxima of ${\bf E}_p$ and $J_\phi$
       and the displacements of their positions  outwards. The impact of $\alpha$
       on the conductivities are ``compressibility"
       dependent: the smaller the $\alpha$ (i) the larger the variation of the maximum of
       $\sigma_\perp$, $\sigma_\parallel$ and $\sigma_\parallel-
       \sigma_\perp$ (when  $M_0$ increases) but (ii) the smaller the maximum
       of $\sigma_\perp$, $\sigma_\parallel$ and $\sigma_\parallel-\sigma_\perp$
       (when $A$ decreases).
 \end{enumerate}

 Qualitatively, except for the conductivity components the impact of the flow on ${\bf E}_p$ and $J_\phi$ are
 independent of ``compressibility". The dependence of the results for $\sigma_\perp$ and
 $\sigma_\parallel$ on ``compressibility" may be due to the fact that
 the incompressible solution (\ref{30}) has finite  flow shear  while the flow of the
 compressible one (\ref{29}) is shearless. Quantitatively, for $\alpha=2$ and
 increase of $M_0$ from 0.1 to 0.5 or decrease of $A$ from -0.001 to -0.006 result in
 percentage variations of all quantities ($\sigma_\parallel$, $\sigma_\perp$, ${\bf E}_p$,
 and $J_\phi$) less that 10\%. Larger variations of these quantities associated with higher
 values of $M_0$ or lower values of $A$, we have found on the basis of solutions (\ref{29})
 and (\ref{30}), most probably overestimate the actual impact of the flow.


 \begin{center}
 {\large\bf Acknowledgement}
 \end{center}
 Part of this work was conducted during a visit of the authors G.P.
 and G.N.T. to the Max-Planck-Institut f\"{u}r Plasmaphysik,
 Garching. The hospitality of that Institute is greatly appreciated.

 %


\begin{thebibliography}{40}
\bibitem{com2} The consequences of the toroidal electric field on the boundary conditions
 are discussed in more detail in Refs. \cite{Th1} and \cite{Th2}.
\bibitem{com1}  To be precise, for incompressible flow when $A$ decreases the position
 of the maximum  first is shifted outwards and  then  inwards
 when $A$ take  small negative values.
\bibitem{com3} For $\alpha=2$ and very small algebraic values of $A$
 i.e. $A\leq -0.08$ it holds that $\sigma_\parallel<\sigma_\perp$.
\end{thebibliography}
 \end{document}